\documentstyle[aps]{revtex}
\begin{document}
\title{Adiabatic Approximation from a Renormalization Group Inspired Method}
\author{C. Contreras$^{a}$\thanks{%
E-mail: ccontrer@fis.utfsm.cl}, J.C. Retamal $^{b}$\thanks{%
E-mail: jretamal@lauca.usach.cl} and L. Vergara$^{b}$\thanks{%
E-mail: lvergara@lauca.usach.cl}}
\address{$^{a}$ Departamento de F\'{\i }sica, Universidad T\'{e}cnica \\
Federico Santa Maria, Casilla 110 - V \\
Valpara\'{\i }so, Chile\\
$^{b}$ Departamento de F\'{\i }sica, Universidad de Santiago \\
de Chile, \\
Casilla 307, Santiago 2, Chile }
\maketitle

\begin{abstract}
We show how a method inspired in renormalization group techniques can be
useful for deriving Hamiltonians in the adiabatic approximation in a
systematic way.
\end{abstract}

%\date{\today}

%\preprint{USACH/00/05, USM-TH-91}

\newpage

Two related renormalization group schemes for Hamiltonians have been
independently proposed in the last years by Wegner \cite{Wegner1} and Glazek
and Wilson \cite{GW} in the context of condensed matter and field theory,
respectively. These nonperturbative methods are useful for diagonalizing or
renormalizing a given Hamiltonian and are based on a continuous sequence of
(infinitesimal) unitary transformations applied to the system Hamiltonian.
The key point in these approaches is that the unitary transformations are
chosen in such a way that off-diagonal matrix elements becomes smaller at
each step of the sequence. The outset of this procedure is a set of flow
equations whose solution yields (in the best case) a (almost) diagonalized
Hamiltonian. These methods have been succesfully applied to various problems
in condensed matter and light front field theory \cite{Otros}.

In this letter we take a look at the origins of this kind of renormalization
transformations, i.e., the work of Fr\"{o}hlich \cite{Frohlich} (see also 
\cite{Dyson}). He gave an explanation for the effective interaction between
electrons in BCS theory of superconductivity by eliminating electron-phonon
interactions using a unitary transformation, which can be viewed as the
''one-step'' formulation of the method mentioned above. We show how
Fr\"{o}hlich method can be used as an approach to the usual adiabatic
approximation used in quantum optics. To that end we will use as a toy model
the Dicke Hamiltonian \cite{Dicke} which describes a collection of two-level
atoms interacting with a quantum field.

In the method of Fr\"{o}hlich a similarity transformation is performed on
the Hamiltonian such that the transformed Hamiltonian has the (troublesome)
off-diagonal elements equal to zero. Since similarity transformations
preserve eigenvalues, the transformed Hamiltonian has the same spectrum as
the original one. Thus the new Hamiltonian is given by

\begin{equation}
H_{U}=UHU^{\dagger }.  \label{dos}
\end{equation}

Let us write the unitary operator as

\begin{equation}
U=e^{-i\Omega }  \label{doss}
\end{equation}

\noindent where $\Omega $ is defined as a perturbative series in a coupling
constant. If we separate the Hamiltonian into its free and interacting parts
we have

\begin{equation}
H_{U}=H_{0}+(H_{I}+i[H_{0},\Omega ])+i[H_{I},\Omega ]-\frac{1}{2}%
[[H_{0},\Omega ],\Omega ]+\cdots  \label{cinco}
\end{equation}

We must look for an operator $\Omega $ such that it eliminates the effect of
the Hamiltonian $H_{I}$ to lowest order, that is, such that the transformed
Hamiltionian is completely diagonal to that order. This means, we must
impose the condition

\begin{equation}
H_{I}+i[H_{0},\Omega ]=0  \label{cincos}
\end{equation}

\noindent which in turn implies that when this condition is satisfied, $%
H_{U} $ can be written as

\begin{equation}
H_{U}=H_{0}+\frac{i}{2}[H_{I},\Omega ]+\cdots  \label{sei}
\end{equation}

To show how the adiabatic approximation is obtained from it we use as a toy
model the Dicke Hamiltonian

\begin{equation}
H=\omega _{0}a^{\dagger }a+\omega _{1}S_{z}+g\ \left( aS_{+}+a^{\dagger
}S_{-}\right)  \label{seis}
\end{equation}

\noindent where $a$ and $a^{\dagger }$ are the usual field operators and the
collective atomic operators satisfy

\begin{equation}
\lbrack S_{+},S_{-}]=2S_{z},\hspace{0.22in}[S_{z},S_{\pm }]=\pm S_{\pm }
\label{siete}
\end{equation}

In the limit of large detuning, $\omega _{1}\gg \omega _{0}$, we can
adiabatically eliminate the transitions among different eigenstates of $%
S_{z}.$ That is, we want to decouple the low frequency modes from the high
frequency modes (this can be done at least perturbatively by diagonalizing
with a suitable unitary operator). By replacing $H_{0}$ and $H_{I}$ into
eqs. (\ref{cinco}) and (\ref{cincos}) we have 
\begin{equation}
\ [\omega _{0}a^{\dagger }a+\omega _{1}S_{z},\Omega ]=ig\ \left\{
aS_{+}+a^{\dagger }S_{-}\right\}  \label{diez}
\end{equation}

\noindent From the structure of this equation and the algebra of the
operators involved, one can easily infer that $\Omega $ must be necessarily
of the form 
\begin{equation}
\Omega =g\left\{ \alpha aS_{+}+\beta a^{\dagger }S_{-}\right\} +\Omega
^{\prime }  \label{diezprim}
\end{equation}
where $\Omega ^{\prime }$ is any operator that commutes with the free
Hamiltonian $H_{0}.$ That is, the solution to eq. (\ref{cincos}) is always
of the form 
\begin{equation}
\Omega =\overline{\Omega }+\Omega ^{\prime }  \label{diiez}
\end{equation}
The fact that $\Omega ^{\prime }$ remains undetermined is harmless. In
effect, we have 
\begin{equation}
H_{U}=e^{-i\Omega ^{\prime }-i\overline{\Omega }}He^{i\Omega ^{\prime }+i%
\overline{\Omega }}  \label{der1}
\end{equation}
but we can always perform a unitary transformation on $H_{U}$ such that{%
\footnote{%
This can be easily proved by following the standard steps one makes when
deriving the Baker-Campbell-Hausdorff or related (e.g Zassenhaus) formulas.}}
\begin{equation}
H_{U}^{\prime }=e^{-i\overline{\Omega }}e^{-i\Omega ^{\prime }}He^{i\Omega
^{\prime }}e^{i\overline{\Omega }}.  \label{der0}
\end{equation}
which means that an unknown unitary transfomation is acting on the original
Hamiltonian $H$: 
\begin{equation}
H^{\prime }=e^{-i\Omega ^{\prime }}He^{i\Omega ^{\prime }}  \label{der2}
\end{equation}
Since a unitary transformation does not change the eigenvalues of the
Hamiltonian, we can always make the choice of working either with $H^{\prime
}$ or $H$. By choosing to work with $H$ is equivalent in practice as
imposing the condition $\Omega ^{\prime }=0.$ We do this in the following.

Therefore, after replacing eq. (\ref{diezprim}) into eq. (\ref{diez}) we get 
\begin{equation}
\Omega =i\frac{g}{\Delta }\left\{ aS_{+}-a^{\dagger }S_{-}\right\}
\label{once}
\end{equation}

\noindent where $\Delta =\omega _{1}-\omega _{0}.$

Also, the ${\cal O}(g^{2})$ contribution

\begin{equation}
H^{(2)}=\frac{i}{2}[H_{I},\Omega ]  \label{doce}
\end{equation}

\noindent is readily obtained from (\ref{seis}) and (\ref{once}): 
\begin{equation}
H^{(2)}=\frac{g^{2}}{\Delta }\left\{ 2S_{z}a^{\dagger }a+S_{+}S_{-}\right\}
\label{trece}
\end{equation}

\noindent which implies that the correct total Hamiltonian to this order in
perturbation theory is 
\begin{equation}
H_{U}=\omega _{0}a^{\dagger }a+\omega _{1}S_{z}+2\frac{g^{2}}{\Delta }%
a^{\dagger }aS_{z}+\frac{g^{2}}{\Delta }\left( S^{2}-S_{z}^{2}+S_{z}\right)
\label{catorce}
\end{equation}

This Hamiltonian can also be derived by using density matrix techniques in
the adiabatic approximation but the calculation is too much involved \cite
{JC}. It has also been derived by using unitary transformations in \cite{JC2}%
, but the way of obtaining it was based on the calculation done in \cite{JC}
and then performing a guess. In that way they obtained 
\begin{equation}
H=\omega _{0}a^{\dagger }a+\omega _{1}S_{z}+2\frac{g^{2}}{\Delta }a^{\dagger
}aS_{z}+\frac{g^{2}}{\Delta }\left( S^{2}-S_{z}^{2}+S_{z}\right) +\frac{g^{2}%
}{2\Delta }\left( S_{+}^{2}+S_{-}^{2}\right)  \label{cator}
\end{equation}

\noindent where the last term had to be discarded by making a further
approximation.

One could have also used a sequence of discrete unitary transformations, as
done e.g. in \cite{Moreno}. After a lengthy calculation one gets 
\begin{equation}
H_{U}=\omega _{0}a^{\dagger }a+\omega _{1}S_{z}+(\frac{g^{2}}{\omega _{1}}+%
\frac{g^{2}\omega _{0}}{\omega _{1}^{2}}+\frac{g^{2}\omega _{0}^{2}}{\omega
_{1}^{3}})\left\{ 2S_{z}a^{\dagger }a+S_{+}S_{-}\right\} +\frac{g\ \omega
_{0}^{2}}{\omega _{1}^{2}}\left\{ aS_{+}+a^{\dagger }S_{-}\right\}
\label{diezb}
\end{equation}

\noindent where the last term must be removed by another unitary
transformation and we recognize in the remaining ${\cal O}(g^{2})$ term part
of the geometric series corresponding to the expansion of $g^{2}/\Delta .$

In this letter we have shown that it is possible to use a renormalization
group inspired method to derive Hamiltonians in the adiabatic, large
detuning, approximation in a straightforward and systematic way, avoiding
the appearance of extra terms that should not be there. It must be stressed
that although the model we used to show the method is simple, the procedure
can be implemented with no difficulties in more realistic models. The
crucial step will always be to find the solution to the operator equations
that appear in the procedure, like eq. (\ref{cincos}) above. This is done by
inspection, which in practice it is not difficult to do.

{\bf Acknowledgements}

This work was supported in part by Fondecyt (Chile) contracts 1970673 and
1990838, DICYT (USACH) under contracts 9931VC and 9934RA, DGIP (USM) under
contract 991121.

\end{document}